\documentclass[shortbibliography,twocolumn,prl,aps,superscriptaddress,showpacs,amsmath,amssymb,floatfix]{revtex4-2}
\usepackage{graphicx}
\usepackage{amssymb}
\usepackage{amsmath}
\usepackage{epsfig}
\usepackage{color}
\usepackage{mathtools}
\usepackage[colorlinks=true, letterpaper=true, pdfstartview=FitV, linkcolor=blue, citecolor=blue, urlcolor=blue]{hyperref}
\usepackage{physics}
\usepackage{siunitx}
\usepackage{bm}
\usepackage{multirow}
\usepackage[dvipsnames]{xcolor}
\definecolor{mypurple}{RGB}{112, 48, 160}
\definecolor{myblue}{RGB}{0, 112, 192}
\usepackage{booktabs}
\begin{document}

\title{Scaling Law for Time-Reversal-Odd Nonlinear Transport}

\author{Yue-Xin Huang}
\affiliation{Department of Physics, City University of Hong Kong, Kowloon, Hong Kong, China}

\author{Cong Xiao}
\email{xiaoziche@gmail.com}
\affiliation{Institute of Applied Physics and Materials Engineering, University of Macau, Taipa, Macau, China}

\author{Shengyuan A. Yang}
\email{yangshengyuan@um.edu.mo}
\affiliation{Institute of Applied Physics and Materials Engineering, University of Macau, Taipa, Macau, China}

\author{Xiao Li}
\email{xiao.li@cityu.edu.hk}
\affiliation{Department of Physics, City University of Hong Kong, Kowloon, Hong Kong, China}

\begin{abstract}
Time-reversal-odd ($\mathcal{T}$-odd) second order nonlinear current response has been theoretically proposed and experimentally confirmed recently. However, the role of disorder scattering in the response, especially whether it contributes to the $\sigma_{xx}$-independent term, has not been clarified. Here, we derive a general scaling law for this effect, which accounts for multiple scattering sources. We show that the nonlinear conductivity is generally a quartic function in $\sigma_{xx}$. The microscopic mechanisms underlying each scaling term are revealed, which include
multiple previously unknown extrinsic processes unique to  $\mathcal{T}$-odd nonlinear transport.
Besides intrinsic contribution, we find extrinsic contributions also enter the zeroth order term, and their values can be comparable to or even larger than the intrinsic one. In addition, cubic and quartic terms must involve skew scattering, and they signal competition between at least two scattering sources.
Our finding reveals the significant role of disorder scattering in $\mathcal{T}$-odd nonlinear transport, and establishes a foundation for analyzing experimental result.
\end{abstract}

\maketitle

The linear anomalous Hall effect (AHE) is a time-reversal odd ($\mathcal{T}$-odd) response, i.e., flipping magnetic moments changes the sign of response, that is only allowed in magnetic materials. The effect has both scattering-irrelevant (intrinsic) and scattering-resulted (extrinsic) origins~\cite{nagaosa2010}.
To investigate the microscopic origins, a standard practice is to plot the measured Hall conductivity versus the longitudinal conductivity $\sigma_{xx}$ and fit the curve with some scaling law~\cite{nagaosa2010,Tian2009scaling,hou2015,Jin2017}. Typically, such an analysis produces a term $\sim (\sigma_{xx})^0\sim \tau^0$ that appears independent of relaxation time $\tau$. This $\sigma^0$-term contains not only the intrinsic AHE~\cite{jungwirth2002,nagaosa2002}, but also extrinsic contributions from so-called side jump and skew scattering mechanisms~\cite{smit_spontaneous_1955,smit_spontaneous_1958,berger_side-jump_1970}. The latter,
which may be categorized as the zeroth order extrinsic (ZOE) contribution, can in principle be comparable to or even larger than the intrinsic one in linear response.

Recently, Hall effects have been extended to nonlinear order. Besides extensive researches on the $\mathcal{T}$-even nonlinear Hall effect~\cite{sodemann_quantum_2015,du2018,facio2018,Sodemann2019,ma2019,kang2019,ortix2019,lee2019strain,Isobe2020,
jiang2020,Shao2020,vanderbilt2020,He2021quantum,Ho2021,Tiwari2021,kumar2021room,Lu2021,ortix2021nonlinear}, it was found that a $\mathcal{T}$-odd nonlinear AHE can exist in magnetic materials and contains an intrinsic contribution~\cite{gao2014,Gao2019,wang2021,liu2021,Zhongbo2023,Huang2023,ma_anomalous_2023,Dimi2023Disorder,Jia2024}. Experiments on antiferromagnetic MnBi$_2$Te$_4$ and Mn$_3$Sn indeed reported a $\sigma^0$-term, which was interpreted as the intrinsic contribution~\cite{Xu2023QM,Gao2023QM,Han2024room,Xu2024electronic}. Nevertheless, thus far, our understanding on the extrinsic contributions to $\mathcal{T}$-odd nonlinear transport is very limited. Especially, there are two outstanding problems. First, the general scaling law for this effect is not known, which severely impedes experimental studies. Second, it is not clear whether its $\sigma^0$-term also contains ZOE contributions or not, which will affect the interpretation of experimental data.


In this work, we solve these two problems. We derive a scaling law for $\mathcal{T}$-odd nonlinear transport [see Eqs.~(\ref{law}), (\ref{ss}) and (\ref{11})]. It shows that plotted against $\sigma_{xx}$, the nonlinear conductivity $\chi_{yxx}$ is generally a quartic function.
The $\sigma^0$-term contains not just the intrinsic contribution, but several ZOE contributions, arising from compositions of side jump and skew scattering, their field corrections, and their combinations with Berry curvature anomalous velocity. The terms beyond zeroth order are all of extrinsic origins. {\color{black} We unveil the mechanisms underlying each scaling term and find multiple new extrinsic processes not present in linear AHE and $\mathcal{T}$-even nonlinear AHE.}
Interestingly, the cubic and the quartic terms must involve skew scattering and arise from competition of at least two scattering sources. We also show that the longitudinal nonlinear conductivity $\chi_{xxx}$ has the similar scaling, in which the ZOE contributions offer a possible explanation for the curious $\sigma^0$-term observed in longitudinal nonlinear signal. The ZOE effects are demonstrated explicitly in a four-band Dirac model, which are found to be comparable to the intrinsic contribution. Our work establishes the framework for comparing theory and experiment and for interpreting experimental data of $\mathcal{T}$-odd nonlinear transport.

\textit{\color{blue} Derivation of the scaling law.} We adopt the semiclassical Boltzmann formalism, which has been successfully applied to obtain scaling laws of linear and $\mathcal{T}$-even nonlinear Hall effects~\cite{sinitsyn_semiclassical_2007,du_disorderinduced_2019}. The electric current density of a system can be expressed as (set $e=\hbar=1$)
\begin{align}
    \bm j=-\sum_l f_l \bm v_l,
    \label{eq-current}
\end{align}
where $l=(n,\bm k)$ is a collective index labeling a Bloch band state,
$\bm v_l$ and $f_l$ are the semiclassical velocity and the distribution function at $l$.

The semiclassical velocity contains three parts
\begin{equation}\label{vv}
  \bm v_l=\partial_{\bm k}\varepsilon_l+\bm E\times \bm{\Omega}_l+\bm v^\text{sj}_l,
\end{equation}
where the first term is the band group velocity, the second is the anomalous velocity due to Berry curvature $\bm{\Omega}_l$~\cite{karplus_hall_1954,chang_berry_1995}, and the last is the side jump velocity due to accumulation of coordinate shifts during scattering~\cite{sinitsyn_disorder_2005,sinitsyn_anomalous_2007}.
To obtain the nonlinear current $\propto E^2$, the first two terms need to include corrections to $E^2$ order, while the last term, as an extrinsic velocity, needs to include $E$-linear correction.

As for $f_l$, it is solved from the Boltzmann equation
\begin{align}
\mathcal{\Hat{D}}_{{E}} f_l= \mathcal{\Hat{I}} f_l
    \label{eq-},
\end{align}
where hat denotes linear operators, $\mathcal{\Hat{D}}_{{E}}=-\bm E \cdot \partial_{\bm k}$ is the field driving term,
and $\mathcal{\Hat{I}}f_l$ is the collision integral. To track extrinsic origins, as shown in Ref. \cite{sinitsyn_semiclassical_2007} (detailed in the Supplemental Material \cite{supp}), it is helpful to separate the collision integral into three parts: $\mathcal{\Hat{I}} =\mathcal{\Hat{I}}_{\text{c}}+\mathcal{\Hat{I}}_{\text{sj}}+\mathcal{\Hat{I}}_{\text{sk}}$, corresponding to the conventional, side jump, and skew scattering processes, respectively.

To obtain the scaling law, the specific form of each term/operator does not concern us here. What matters is its scaling
with respect to field $E$ and disorder strength $V$. We shall use the notation $Q^{(i,j)}$ to indicate a quantity $Q\propto E^i V^{-j}$. For example,
$
  \mathcal{\Hat{I}}_{\text{c}} f_l=-\sum_{l'}\omega_{l'l}(f_l-f_{l'}),
$
where $\omega_{l'l}$ is the scattering rate from $l$ to $l'$. Assuming weak disorders, the leading
contribution to $\mathcal{\Hat{I}}_{\text{c}}$ is at $V^2$ order, hence we may write
\begin{equation}\label{csc}
  \mathcal{\Hat{I}}_c=\mathcal{\Hat{I}}_c^{(0,-2)}.
\end{equation}
The similar analysis can be readily done for other terms. It is worth noting that to capture $E^2$ response, especially the ZOE contribution, one must include $E$ field corrections to operators $\mathcal{\Hat{I}}_\text{sj}$ and $\mathcal{\Hat{I}}_\text{sk}$ ($\mathcal{\Hat{I}}_\text{sj}$ is already linear in $E$~\cite{Xiao2019NLHE}, thus its correction is of $E^2$ order). Hence, their scaling behaviors are described by
\begin{align}
  \mathcal{\Hat{I}}_\text{sj}&=\mathcal{\Hat{I}}_\text{sj}^{(1,-2)}+\mathcal{\Hat{I}}_\text{sj}^{(2,-2)}, \\
  \mathcal{\Hat{I}}_\text{sk}&=\mathcal{\Hat{I}}_\text{sk}^{(0,-3)}+\mathcal{\Hat{I}}_\text{sk}^{(1,-3)}+\mathcal{\Hat{I}}_\text{sk}^{(0,-4)}+\mathcal{\Hat{I}}_\text{sk}^{(1,-4)}.
  \label{sk}
\end{align}
Note that in $\mathcal{\Hat{I}}_\text{sk}$, besides the leading order ($V^3$) terms from third-order asymmetric scattering, it is necessary to also include fourth order processes [the latter two terms in (\ref{sk})] which are known as intrinsic skew scattering \cite{sinitsyn_anomalous_2007}. Processes with even
higher order do not bring in new scaling behavior, but just renormalize the existing terms.

\renewcommand{\arraystretch}{1.8}
\begin{table*}
    \centering
        \caption{\color{black}Physical interpretation of the terms in scaling law (\ref{law}). Here, A, E, SJ, SK, and ISK denote
        anomalous velocity, $E$-field correction, side jump, conventional (third-order) skew scattering, and intrinsic skew scattering, respectively. The third column gives the concrete expressions of the contributions to 
        $\mathcal{T}$-odd nonlinear response current. In these expressions, $\{\cdots\}$ means the anti-commutator, $\bm v^\mathrm{b}\equiv \partial_{\bm k}\varepsilon$ is the band group velocity, and 
        $\delta^{ E} \bm v^\mathrm{sj}$ is the $E$-field correction of side jump velocity.
        }
    \begin{tabular}{p{0.11\linewidth}<{\centering}p{0.17\linewidth}<{\centering}p{0.695\linewidth}<{\centering}}
        \hline\hline
        Term in (\ref{law}) & Mechanism & Expression
        \\ \hline
        $D$ & Drude &
        $-\tau^2\sum_l \hat D_E\hat D_E f_0$
        \\
        $c_i$ & A-SJ, E-SJ &
        $-\tau\sum_l \left[ \left( \bm E\times \bm \Omega \right) \hat{\mathcal I}_\mathrm{sj}^{(1,-2)}-\delta^{ E}\bm v^\mathrm{sj}\hat{D}_E+\bm v^\mathrm{b} \hat{\mathcal I}_\mathrm{sj}^{(2,-2)} \right]f_0$
        \\
        $c_{ij}$ & SJ-SJ, A-ISK, E-ISK &
        $ -\tau^2\sum_l
        \left[
            \bm v^\mathrm{b} \hat{\mathcal I}_\mathrm{sj}^{(1,-2)}\hat{\mathcal I}_\mathrm{sj}^{(1,-2)}
            - \bm v^\mathrm{sj}\acomm{\hat D_E}{\hat{\mathcal I}_\mathrm{sj}^{(1,-2)}}
            -(\bm E\times \boldsymbol\Omega) \hat{\mathcal I}_\mathrm{sk}^{(0,-4)}\hat D_E
            -\bm v^\mathrm{b} \hat{\mathcal I}_\mathrm{sk}^{(1,-4)}\hat D_E
        \right]f_0$
        \\
        $c_{ijk}$ & SJ-ISK &
        $ -\tau^3\sum_l
        \left[
            \bm v^\mathrm{sj}\acomm{\hat D_E}{\hat{\mathcal I}_\mathrm{sk}^{(0,-4)}}\hat D_E
            -\bm v^\mathrm{b} \acomm{\hat D_E}{\hat{\mathcal I}_\mathrm{sk}^{(0,-4)}} \hat{\mathcal I}_\mathrm{sj}^{(1,-2)}
            -\bm v^\mathrm{b} \acomm{\hat{\mathcal I}_\mathrm{sk}^{(0,-4)}}{\hat{\mathcal I}_\mathrm{sj}^{(1,-2)}} \hat D_E
        \right]f_0 $
        \\
        $c_{ijk\ell}$ & ISK-ISK &
        $-\tau^4\sum_l \bm v^\mathrm{b}
        \left[
            \hat D_E\hat{\mathcal I}_\mathrm{sk}^{(0,-4)} \hat{\mathcal I}_\mathrm{sk}^{(0,-4)}
            +\hat{\mathcal I}_\mathrm{sk}^{(0,-4)}\hat D_E \hat{\mathcal I}_\mathrm{sk}^{(0,-4)}
            +\hat{\mathcal I}_\mathrm{sk}^{(0,-4)} \hat{\mathcal I}_\mathrm{sk}^{(0,-4)}\hat D_E
        \right]f_0$
        \\
        $d_i$ & A-SK, E-SK &
        $\tau^2\sum_l\left[ (\bm E\times \boldsymbol\Omega)\hat{\mathcal I}_\mathrm{sk}^{(0,-3)}\hat D_E+\bm v^\mathrm{b}\hat{\mathcal I}_\mathrm{sk}^{(1,-3)}\hat D_E \right]f_0$
        \\
        $d_{ij}$ & SJ-SK &
        $-\tau^3\sum_l
        \left[
            \bm v^\mathrm{sj}\acomm{\hat D_E}{\hat{\mathcal I}_\mathrm{sk}^{(0,-3)}}\hat D_E
            +\bm v^\mathrm{b} \acomm{\hat D_E}{\hat{\mathcal I}_\mathrm{sk}^{(0,-3)}} \hat{\mathcal I}_\mathrm{sj}^{(1,-2)}
            +\bm v^\mathrm{b} \acomm{\hat{\mathcal I}_\mathrm{sk}^{(0,-3)}}{\hat{\mathcal I}_\mathrm{sj}^{(1,-2)}} \hat D_E
        \right]f_0$
        \\
        $d_{ijk}$ & ISK-SK &
        $-\tau^4\sum_l\bm v^\mathrm{b}
        \left[
            \hat D_E\acomm{\hat{\mathcal I}_\mathrm{sk}^{(0,-3)}}{\hat{\mathcal I}_\mathrm{sk}^{(0,-4)}}
            +\hat{\mathcal I}_\mathrm{sk}^{(0,-3)}\acomm{\hat D_E}{\hat{\mathcal I}_\mathrm{sk}^{(0,-4)}}
            +\hat{\mathcal I}_\mathrm{sk}^{(0,-4)}\acomm{\hat D_E}{\hat{\mathcal I}_\mathrm{sk}^{(0,-3)}}
        \right]f_0$
        \\
        $\mathcal{D}_{ij}$ & SK-SK &
        $-\tau^4\sum_l \bm v^\mathrm{b}
        \left[
            \hat D_E\hat{\mathcal I}_\mathrm{sk}^{(0,-3)}\hat{\mathcal I}_\mathrm{sk}^{(0,-3)}
            +\hat{\mathcal I}_\mathrm{sk}^{(0,-3)}\hat D_E \hat{\mathcal I}_\mathrm{sk}^{(0,-3)}
            +\hat{\mathcal I}_\mathrm{sk}^{(0,-3)} \hat{\mathcal I}_\mathrm{sk}^{(0,-3)}\hat D_E
        \right]\hat D_E f_0$
        \\ \hline\hline
    \end{tabular}
    \label{expression}
\end{table*}

To solve Eq.~(\ref{eq-}), we decompose $f$ as a sum
\begin{equation}\label{ff}
  f=f^0+\sum_{i=1,2;j}f^{(i,j)},
\end{equation}
whereb $f^0$ is the equilibrium Fermi-Dirac distribution, and the off-equilibrium part is written using our notation above.
Here, $i$ is restricted to 1 and 2 because we are seeking $E^2$ response. For a fixed $i$, the cutoff of $j$ is given by the one that allows $f$ to be solved consistently from Eq.~(\ref{eq-}), which can be shown to be $2i$.
Substituting (\ref{csc}-\ref{ff}) into (\ref{eq-}), one can decompose the kinetic equation
into a set of linear equations according to the $(E,V^{-1})$ order of terms, as presented in Supplemental Material \cite{supp}.

To obtain a general scaling law, we take into account the fact that a real system usually has more than one sources of scattering. Assuming there is no correlation between different scattering sources, the Matthiessen's rule holds, i.e., $1/\tau=\sum_i (1/\tau_i)$, or equivalently, $\rho_{xx}=\sum_i\rho_i$, where $\tau_i$ and $\rho_i$ are relaxation time and partial resistivity for the $i$-th type scattering.
The scaling behavior of each $f^{(i,j)}$ in terms of the $\rho$'s is readily obtained from solving the kinetic equation, as detailed in Supplemental Material~\cite{supp}.
We checked that this approach reproduces the scaling laws for both linear AHE~\cite{hou2015,Jin2017} and $\mathcal{T}$-even nonlinear AHE~\cite{du_disorderinduced_2019} obtained previously.

Here, we extract and focus on the $\mathcal{T}$-odd contributions at $E^2$ order.
For the transverse nonlinear response, we arrive at the following general scaling relation
\begin{align}
     \chi_{yxx}=&\ \frac{D}{\rho_{xx}^2}+c^\text{int}+\sum_i c_i\frac{\rho_i }{\rho_{xx}}+\sum_{ij}c_{ij}\frac{\rho_i\rho_j}{\rho_{xx}^2} \nonumber\\
  &+\sum_{ijk}c_{ijk}\frac{\rho_i\rho_j\rho_k}{\rho_{xx}^3}+
\sum_{ijk\ell}c_{ijk\ell}\frac{\rho_i\rho_j\rho_k\rho_\ell}{\rho_{xx}^4} \nonumber\\
  &+\sum_{i\in S}\bigg(d_i+\sum_j d_{ij}\frac{\rho_j}{\rho_{xx}}+\sum_{jk} d_{ijk}\frac{\rho_j\rho_k}{\rho_{xx}^2}\nonumber\\
  &\qquad+ \sum_{j\in S} \mathcal{D}_{ij}\frac{\rho_j}{\rho_{xx}^2}\bigg)\frac{\rho_i}{\rho_{xx}^2} \label{law},
\end{align}
where all $c$'s and $d$'s, $D$, and $\mathcal{D}$ are scaling parameters independent of each disorder concentration, and $i\in S$ means it counts only static impurities. One recognizes that $c^\text{int}$ is the intrinsic contribution
predicted in Refs.~\cite{gao2014,wang2021,liu2021}, and $D/\rho_{xx}^2$ term is a nonlinear Drude like response~\cite{Souza2022,Liu2022PRL}. The physical meaning of other scaling terms will be analyzed below.

\textit{\color{blue} Interpretation of the scaling law.} {The scaling law for $\mathcal{T}$-odd nonlinear transport exhibits features distinct from linear and $\mathcal{T}$-even nonlinear AHEs. 

{\color{black} First of all, let us clarify the microscopic mechanisms underlying each scaling term in Eq.~(\ref{law}). This is made particularly transparent in our approach, by keeping track of scattering operators involved in each contribution.
The results are summarized in Table~\ref{expression}. The second column lists the physical origins, and the third column
gives the concrete expressions. For example, $c_i$ is from the composition of anomalous velocity and side jump (A-SJ) and 
the $E$-field correction of side jump (E-SJ). $c_{ij}$ has three origins: composition of two side jumps (SJ-SJ), composition of anomalous velocity with intrinsic skew scattering (A-ISK), and $E$-field correction of intrinsic skew scattering (E-ISK). 
The $d$ and $\mathcal{D}$ terms all involve third-order skew scattering, denoted as SK in the acronyms. Interpretations of other terms can be easily read off from the table. It is worth noting that, here, multiple terms involve compositions of scattering mechanisms, such as SJ-SJ, SJ-SK, and etc. This does not occur in linear AHE or $\mathcal{T}$-even nonlinear AHE, where each scattering process act only once and independently.}

As mentioned, in experiment, the scaling plot is often made in terms of conductivities. Here, we specialize to the mostly encountered case, where the system contains two major competing scattering sources: one static (e.g., impurities, defects, and boundary roughness) and one dynamic from phonon, labeled by $i=0$ and $1$, respectively~\cite{Tian2009scaling,hou2015,Jin2017}.
Using $\sigma_{xx}\approx 1/\rho_{xx}$ and $\sigma_0\equiv\sigma_{xx}(T=0)$ denoting the longitudinal conductivity at low temperature, we find
%
%
%
\begin{align}
     \chi_{yxx}=& D\sigma_{xx}^2+ \lambda_0+\lambda_1\frac{\sigma_{xx}}{\sigma_0}+\lambda_2\frac{\sigma_{xx}^2}{\sigma_0^2}+\lambda_3\frac{\sigma_{xx}^3}{\sigma_0^3}
  +\lambda_4\frac{\sigma^4_{xx}}{\sigma_0^4} \nonumber\\
  &+\eta_2\frac{\sigma^2_{xx}}{\sigma_0}+\eta_3\frac{\sigma^3_{xx}}{\sigma_0^2}+\eta_4\frac{\sigma^4_{xx}}{\sigma_0^3}+\gamma_{4}\frac{\sigma^4_{xx}}{\sigma_0^2}. \label{ss}
\end{align}
Here, $\gamma_{4}=\mathcal{D}_{00}$, and the coefficients $\lambda$'s ($\eta$'s) here can be expressed using the $c$'s ($d$'s) in (\ref{law}). For example,
\begin{equation}\label{t0}
  \lambda_0=c^\text{int}+c_{1}+c_{11}+c_{111}+c_{1111},
\end{equation}
and other expressions can be found in \cite{supp}. From (\ref{ss}), one observes that $\chi_{yxx}$ is generally a quartic function of $\sigma_{xx}$. This is distinct from the scaling relations for linear AHE and $\mathcal{T}$-even nonlinear AHE, which are quadratic and cubic functions in $\sigma_{xx}$, respectively. 

Importantly, $\lambda_0$ in the scaling law (\ref{ss}) is just the $\sigma^0$-term, which is a distinct feature for $\mathcal{T}$-odd transport and not present in $\mathcal{T}$-even response. From its expression in (\ref{t0}), one can clearly see that besides the intrinsic contribution, there indeed exist ZOE contributions. {\color{black} From Table~\ref{expression}, these ZOE contributions originate from multiple mechanisms, including A-SJ, E-SJ, SJ-SJ, A-ISK, E-ISK, SJ-ISK, and ISK-ISK.}

\renewcommand{\arraystretch}{1.3}
\begin{table*}
  \caption{\color{black}Comparison of scaling laws for linear, $\mathcal{T}$-even nonlinear, and $\mathcal{T}$-odd nonlinear AHEs, in the presence of two types of disorder (one static and one dynamic). Here, BC, BCD, and BCP represent Berry curvature, Berry curvature dipole, and Berry connection polarizability contributions, respectively. We drop the subscript `$xx$' in $\sigma$'s for simple notations. The ZOE contributions are highlighted with red color.
  } \label{comparison}
    \begin{centering}
      \begin{tabular}{p{0.07\linewidth}<{\centering}
        p{0.05\linewidth}<{\centering}
        p{0.16\linewidth}<{\centering}
        p{0.20\linewidth}<{\centering}
        p{0.48\linewidth}<{\centering}}
        \hline\hline
        \multicolumn{2}{c}{Scaling terms}  & Linear AHE & $\mathcal{T}$-even nonlinear AHE & $\mathcal{T}$-odd nonlinear AHE \\
        \hline
        \specialrule{0em}{0pt}{3.5pt}
        $\sigma^0$ &  & BC, SJ, ISK & & {BCP}, \color{red}{A-SJ, E-SJ, A-ISK, E-ISK, SJ-SJ, SJ-ISK, ISK-ISK}
        \vspace{6pt} \\
        \multirow{2}{*}{$\sigma^1$} & ${\sigma}/{\sigma_0}$ & SJ, ISK & & A-SJ, E-SJ, A-ISK, E-ISK, SJ-SJ, SJ-ISK, ISK-ISK \\
        &$\sigma/\sigma_0^0$ & & BCD, SJ, ISK & \quad
        \vspace{6pt}\\
        \multirow{3}{*}{$\sigma^2$} & ${\sigma^2}/{\sigma_0^2}$ & ISK & & A-ISK, E-ISK, SJ-SJ, SJ-ISK, ISK-ISK \\
        & ${\sigma^2}/{\sigma_0}$ & {SK} &	SJ, ISK &	{A-SK, E-SK, SJ-SK, ISK-SK} \\
        & $\sigma^2/\sigma_0^0$ & & & {nonlinear Drude}
        \vspace{6pt}\\
        \multirow{3}{*}{$\sigma^3$} & ${\sigma^3}/{\sigma_0^3}$ & & & SJ-ISK, ISK-ISK \\
        & ${\sigma^3}/{\sigma_0^2}$ & & ISK &	{SJ-SK, ISK-SK} \\
        & ${\sigma^3}/{\sigma_0}$ & & {SK} & \quad
        \vspace{6pt} \\
        \multirow{3}{*}{$\sigma^4$} & ${\sigma^4}/{\sigma_0^4}$ & & & ISK-ISK \\
        &${\sigma^4}/{\sigma_0^3}$ & & & {ISK-SK} \\
        &${\sigma^4}/{\sigma_0^2}$ & & & {SK-SK}
        \vspace{3pt}\\
        \hline\hline

      \end{tabular}
    \end{centering}
\end{table*}

{\color{black} The analysis can be extended to other scaling terms in Eq.~(\ref{ss}). In Table~\ref{comparison}, we present the interpretation of each term and make a comparison with the scaling laws of linear AHE and $\mathcal{T}$-even nonlinear AHE.
One observes that both linear and $\mathcal{T}$-even AHE involve four scaling terms, and the highest scaling power in $\sigma_{xx}$ is two orders higher than the lowest one (2 and 0 in linear AHE; 3 and 1 in $\mathcal{T}$-even nonlinear AHE). This similarity is rooted in the common physical mechanisms involved, as can be seen from the second and the third columns of Table~\ref{comparison}.
In contrast, the content of $\mathcal{T}$-odd nonlinear response is much richer and distinctive. It contains ten scaling terms, and the highest scaling power is four orders higher than the lowest (4 and 0). Linear AHE and $\mathcal{T}$-even nonlinear AHE have 3 extrinsic mechanisms (SJ, SK, and ISK), whereas $\mathcal{T}$-odd nonlinear transport involves 
13 extrinsic mechanisms. Among them, nonlinear Drude and A-SK contributions were proposed in recent works \cite{Souza2022,Liu2022PRL,ma_anomalous_2023}, and all others have not been studied before.}

Regarding Table~\ref{comparison}, we have three additional remarks. First, the $\sigma^4$-term is present in 
$\mathcal{T}$-odd nonlinear transport but not in linear and $\mathcal{T}$-even nonlinear responses. Second, 
although some scaling term (e.g., the ${\sigma^3}/{\sigma_0^2}$ term) in $\mathcal{T}$-odd nonlinear transport appears also in linear or $\mathcal{T}$-even nonlinear responses, their underlying mechanisms are different.
Third, one notes that $\sigma^3$- and $\sigma^4$-terms in $\mathcal{T}$-odd nonlinear transport 
must originate from compositions of two scattering processes, and at least one of them must be
skew scattering (SK or ISK).

We have also investigated the longitudinal $\mathcal{T}$-odd response $\chi_{xxx}$. The obtained scaling law for $\chi_{xxx}$ has the same form as (\ref{ss})~\cite{supp}, with the only difference that there is no $c^\text{int}$ term in $\lambda_0$~\cite{gao2014,Gao2019,wang2021,liu2021}, consistent with the expectation that intrinsic contribution should not appear in a dissipative current~\cite{Xiao2024definition}.
Importantly, one finds that $\chi_{xxx}$ for a $\mathcal{T}$-broken system can also have a $\sigma^0$-term in the scaling relation, from ZOE contributions.

Finally, we note that even in the simple case of a single type of disorder, our scaling law uncovers previously unknown interesting features.
In this case, Eq.~(\ref{law}) reduces to (label the disorder here by `1')
\begin{equation}\label{11}
\chi_{yxx}=(c^\text{int}+c_{1}+c_{11}+c_{111}+c_{1111})+A\sigma_{xx}+B\sigma_{xx}^2,
\end{equation}
where $A=d_1+d_{11}+d_{111}$ and $B=D+\mathcal{D}_{11}$.
We have two observations. First, compared with (\ref{ss}), one can see that
the appearance of sizable cubic and/or quartic terms in scaling indicates the presence of at least two competing scattering sources.
{\color{black} Second, the $\sigma^0$-term here still contains ZOE and has the same interpretation as in (\ref{ss}), however, the $\sigma^1$- and $\sigma^2$-terms
acquire a different meaning. The $\sigma^1$-term here originates from A-SK, E-SK, SJ-SK, and ISK-SK mechanisms; while
the $\sigma^2$-term comes from Drude and SK-SK mechanisms.}

\textit{\color{blue} ZOE terms in Dirac model.}
To illustrate the features of ZOE contributions,
we explicitly evaluate them in the four-band Dirac model, which is the minimal model that hosts $\mathcal{PT}$-symmetry and $\mathcal T$-odd nonlinear transport~\cite{wang2021,liu2021}. It can be expressed as
\begin{align}
   \mathcal{H}= w k_y +v(k_x\sigma_x+k_y\sigma_y)+\Delta \sigma_z s_z
    \label{eq-H2d},
\end{align}
where $\sigma_i$'s and $s_i$'s are the Pauli matrices for orbital and spin degrees of freedom, respectively; $w$, $v$, and $\Delta$ are model parameters. The tilt $w$ is needed to break $C_{2z}$ symmetry to allow in-plane nonlinear response.
For the disorder part, we consider a single type of disorder, and adopt the white noise averages $\langle V\rangle_c =0$ and $\langle V^2\rangle_c> 0$ with all higher-order correlations vanishing. Thus, $A=\mathcal{D}_{11}=0$ in Eq.~(\ref{11}), and we focus on the ZOE terms.

\begin{figure}[b]
    \centering
    \includegraphics[width=0.49\textwidth]{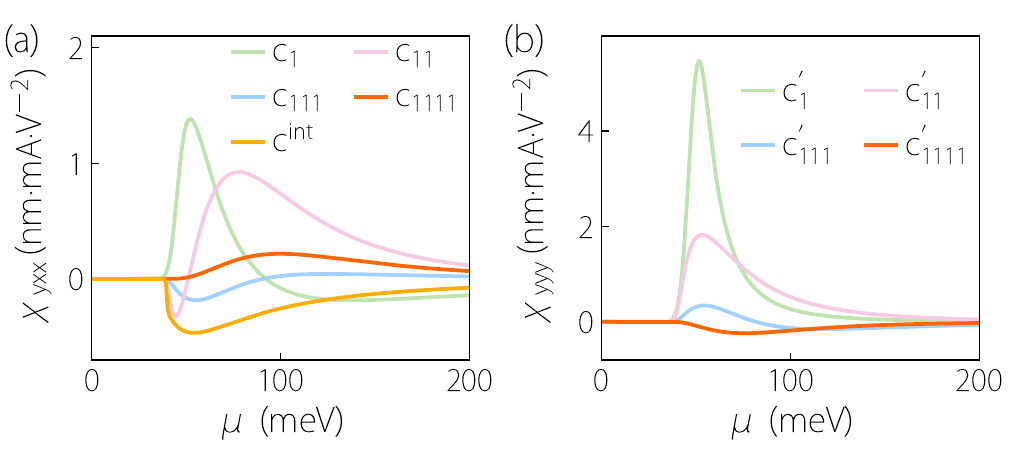}
    \caption{Contributions to the $\sigma^0$-term of (a) transverse and (b) longitudinal nonlinear conductivities for the Dirac model. In the calculation, we take $v=\SI{1e6}{m/s}$, $\Delta=\SI{40}{meV}$, and $w/v=0.2$.}
    \label{fig-Transverse}
\end{figure}

For this specific model, the ZOE contributions can be explicitly evaluated using our approach. The full analytic expressions are given in \cite{supp}. Here, we plot the results in Fig.~\ref{fig-Transverse}.
Figure~\ref{fig-Transverse}(a) shows the variation of intrinsic and ZOE terms in $\chi_{yxx}$ versus chemical potential $\mu$. One observes that the
ZOE terms, like the intrinsic one, appear to be enhanced near the Dirac band edge and decrease when moving to higher energies.
Compared with intrinsic contribution, their magnitudes can be comparable or even larger, and they may have a different sign. In Fig.~\ref{fig-Transverse}(a), taking into account the ZOE contribution leads to a sign change of the $\sigma^0$-term. In Fig.~\ref{fig-Transverse}(b), we plot the contributions to the $\sigma^0$-term for $\chi_{yyy}$. The qualitative behavior is similar to Fig.~\ref{fig-Transverse}(a), confirming that the ZOE contributions can appear as $\sigma^0$-term in longitudinal nonlinear conductivity.

\textit{\color{blue}Discussion.} We have established a scaling law for $\mathcal{T}$-odd nonlinear transport, and through the process, clarified the mechanisms of each scaling term and the presence of ZOE contributions to both transverse and longitudinal responses. The general scaling law in Eq.~(\ref{law}) and the specialized forms in
Eq.~(\ref{ss}) and Eq.~(\ref{11}) provide the foundation for
analyzing and interpreting experimental results.

For measurement on magnetic systems, usually both $\mathcal{T}$-odd and $\mathcal{T}$-even responses are present. As mentioned, the scaling for $\mathcal{T}$-even response has been studied~\cite{du_disorderinduced_2019} and the terms are also recovered in our derivation.
Comparing the two, we note that the $\sigma^0$-term and the $\sigma^4$-term for $\mathcal{T}$-odd response do not exist in $\mathcal{T}$-even response.
According to their definitions, the two responses can be separated by comparing the experimental measurements on the two magnetic configurations connected by $\mathcal{T}$. Certain magnetic space group symmetries may also help to eliminate one in favor of the other.
For example, $\mathcal{PT}$ or $C_{2z}\mathcal{T}$ symmetry ($\mathcal{P}$ is the inversion, and $C_{2z}$ the twofold rotation) can suppress the $\mathcal{T}$-even nonlinear response in the plane, then only the $\mathcal{T}$-odd one exists.

In our derivation, the result is given for the total transverse nonlinear response $\chi_{yxx}$. Sometimes, it may be useful to separate $\chi_{yxx}$ into dissipationless and dissipative parts~\cite{Sodemann2019,liu2022}: $\chi_{yxx}=\chi_{[yx]x}+\chi_{(yx)x}$, where the square (round) bracket means antisymmetrization (symmetrization) of indices. We find that both parts follow the same scaling law, except that the intrinsic contribution enters only the dissipationless part (whereas ZOE enters both). The two parts may have different behaviors under crystalline symmetry. For instance, under $C_{3z}$ symmetry, $\chi_{[yx]x}$ is suppressed, whereas $\chi_{(yx)x}$ is allowed. For the Dirac model here, our calculation shows that both parts are present and have comparable magnitudes~\cite{supp}. Experimentally, the two parts can be separated via a sum-frequency generation technique~\cite{Xu2023QM,Gao2024antiferromagnetic}.

\textit{\color{blue}Note added.} {\color{black} During the review process of our paper, two related works~\cite{Mehraeen2024,Lu2024nonlinear} appeared, which support our result in the overlapping part.}

\bibliography{ref}
\bibliographystyle{apsrev4-2}

\end{document}